\documentclass[twocolumn,prb,aps,showpacs,superscriptaddress]{revtex4}

\usepackage{graphicx}

\def\be{\begin{equation}}
\def\ee{\end{equation}}
\def\ba{\begin{eqnarray}}
\def\ea{\end{eqnarray}}

\def\LSCO{La$_{2-x}$Sr$_x$CuO$_4$ }

\def\YBCO{YBa$_2$Cu$_3$O$_{7-\delta}$ }
\def\124{YBa$_2$Cu$_4$O$_8$ }

\def\C60{A$_x$C$_{60}$ }

\def\LBCO{La$_{2-x}$Ba$_{x}$CuO$_4$ }

\def\18{\frac{1}{8}}

\begin{document}

\title{Spatial modulations of mid-gap states in
(001)La$_{1.88}$Sr$_{0.12}$CuO$_4$ films - indications for
anti-phase ordering of the \emph{d}-wave order parameter}

\author{Ofer Yuli} \email{ofer.yuli@mail.huji.ac.il}
\affiliation{Racah Institute of Physics, The Hebrew University of
Jerusalem, Jerusalem 91904, Israel}

\author{Itay Asulin}
\affiliation{Racah Institute of Physics, The Hebrew University of
Jerusalem, Jerusalem 91904, Israel}

\author{Gad Koren}
\affiliation{Department of Physics, Technion - Israel Institute of
Technology, Haifa 32000, Israel}


\author{Oded Millo}
\affiliation{Racah Institute of Physics, The Hebrew University of
Jerusalem, Jerusalem 91904, Israel}

\begin{abstract}

Using scanning tunneling spectroscopy we have investigated the
spatial evolution of the anomalous \emph{c}-axis zero bias
conductance peak, discovered in a previous study by our group, in
epitaxial La$_{1.88}$Sr$_{0.12}$CuO$_4$ thin films. We found an
anisotropic spatial dependence of the corresponding low-energy
density of states which complies with the predicted spectral
features of an anti-phase ordering of the \emph{d}-wave order
parameter within the \emph{ab}-plane. Such an ordering was
recently suggested to account for the 1/8 anomaly in the high
temperature superconductors and the dynamical layer decoupling
recently reported to occur in the transport studies of
La$_{15/8}$Ba$_{1/8}$CuO$_4$.

\end{abstract}


\maketitle

\section{INTRODUCTION}\label{sec:intro}

The family of lanthanum-based high temperature superconductors,
La214, exhibits an anomalous drop of the transition temperature,
$T_c$, at the x = 1/8 doping level, an effect also known as the
1/8 anomaly. In \LSCO (LSCO) $T_c$ drops at x = 1/8 by about 30\%
with respect to the value of the 'unperturbed' superconductor
dome, \cite{Matsuzaki, Sato2} whereas in \LBCO the effect is much
more prominent and $T_c$ reduces to almost zero.\cite{Mood} At the
same doping level, an anomalously small width of the peak
momentum, associated with the electronic stripe phase was measured
by neutron scattering on LSCO,\cite{Yamada} which in addition,
exhibited a commensurate ordering of the stripes with the
underlying lattice.\cite{Yamada,Tran} The coincidence of the two
phenomena at x = 1/8 has led to the conjecture that the stripe
order competes with superconductivity, and their strong
interaction is responsible for this anomaly.\cite{Himeda, Orgad
PRB} This connection, however, has not been fully established yet,
neither theoretically nor experimentally.

Recently, the transport properties of the stripe phase were
examined in La$_{15/8}$Ba$_{1/8}$CuO$_4$ single crystals by Li
\emph{et al.}.\cite{Li} Below the charge and spin ordering
temperatures, a 2D superconducting Berezinskii-Kosterlitz-Thouless
(BKT) \cite{B,KT} transition was identified. The extracted
$T_{BKT}$ exceeded the bulk $T_c$ ($<$4 K), implying that for the
$T_c < T < T_{BKT}$ temperature range, superconductivity is
confined to two-dimensional planes with negligible inter-plane
coupling, a behavior which was confirmed in a recent
photo-emission study of La$_{15/8}$Ba$_{1/8}$CuO$_4$.\cite{He}
Shortly after, Berg \emph{et al.} \cite{Berg} suggested that an
anti-phase ordering of the order parameter within each CuO$_2$
plane accounts for the above two transition temperatures reported
by Li and coworkers. Apparently, this type of order parameter
modulation within the \emph{ab}-plane suppresses inter-layer
Josephson coupling, while in-plane superconductivity survives.

The local density of states (LDOS) of a phase-biased junction
comprising two \emph{d}-wave superconductors was calculated by
Tanaka and Kashiwaya \cite{TK LDOS} who found that for a $\pi$
phase difference, zero-energy Andreev bound states (ABSs) form at
the interface. Such a junction is analogous to a domain wall where
the order parameter undergoes a $\pi$ phase shift ($\pi$DW),
described in the anti-phase ordering model put forward by Berg
\emph{et al.}.\cite{Berg} The spatial evolution of the LDOS as a
function of distance from a $\pi$DW, as well as the corresponding
effects of the pairing amplitude, were calculated by Yang \emph{et
al.}.\cite{Sigrist} In the weak pairing regime of this
calculation, $\Delta << E_F$, the corresponding zero-bias
conductance peak (ZBCP) appears in the LDOS at the vicinity of the
$\pi$DW. A higher pairing amplitude, on the other hand, is
predicted to induce an imbalanced splitting of the ZBCP at small
distances from the DW due to the formation of a one-dimensional
band of propagating ABSs along the $\pi$DW. For both pairing
amplitudes studied, the spectral weight shifts from zero energy
toward finite energies corresponding to the bulk
superconductor-gap edges, as a function of distance from the DW.
Away from the $\pi$DW, at the center of the domain, the LDOS was
predicted to exhibit a suppressed peak at zero energy alongside
pronounced peaks at the bulk gap edges. In contrast, for the more
realistic scenario of an array of $\pi$DWs with the conventional
4a$_0$ periodicity of the stripe ordering (where a$_0$ is the LSCO
lattice constant), the calculated LDOS exhibited a gap centered at
the chemical potential instead of the ZBCP. It was argued that the
overlap of neighboring ABSs couples adjacent $\pi$DWs which yields
a total quenching of the ZBCP. \cite{Shirit, Sigrist}

In spite of the compatibility of the anti-phase model \cite{Berg}
with the results of Li \emph{et al.},\cite{Li} the existence of an
anti-phase ordering has not yet been established experimentally.
An indication for the existence of an isolated $\pi$DW phase was
found in a previous study of our group,\cite{Yuli PRB} albeit at
the time this connection was not recognized. There we examined the
LDOS near the x = 1/8 doping level on \emph{c}-axis
La$_{1.88}$Sr$_{0.12}$CuO$_4$ films, using scanning tunneling
spectroscopy. Although a small fraction of the data comprised the
expected \emph{c}-axis V-shaped tunneling gap, we ascertained that
the dominant spectral feature in the LDOS was a ZBCP. This
surprising phenomenon was uniquely found in films with $x = 0.12$
and replaced the V-shaped gap found in all other (001)\LSCO
samples we examined with $x \neq 0.12$. Noting that the ZBCP is a
hallmark feature of \emph{d}-wave superconductivity,\cite{Hu,TK}
appearing in tunneling spectra measured on nodal or any in-plane
orientated surface [except the anti-nodal (100) surface], we
looked for evidence for extraneous (\emph{e.g.} faceting induced)
in-plane tunneling in our results, but to no avail. Within the
sensitivity of our x-ray diffraction measurements, no indication
for any in-plane oriented surface was found. On the other hand,
sharp Bragg peaks, corresponding to a highly ordered \emph{c}-axis
phase, were detected in all of our films. In addition, facet
tunneling effects (i.e. coupling to nodal surfaces residing at the
side facets of the \emph{c}-axis grains) were also ruled out,
since our data revealed a spatial continuity of the ZBCP feature
\emph{on top} of the \emph{c}-axis grains. Moreover, a finite ZBCP
amplitude was detectable for all temperatures below $T_c$,
contrary to the commonly reported disappearance of the nodal ABSs
at temperatures well below $T_c$.\cite{Dagan, Greene2} Thus,
despite the similar spectral signature of our anomalous
\emph{c-}axis ZBCPs to tunneling spectra portraying conventional
nodal surface ABSs, we concluded in Ref. \onlinecite{Yuli PRB}
that our findings have a different, yet unresolved, origin.

Motivated by the newly proposed anti-phase ordering at x =
1/8,\cite{Berg} and its possible relation to our previously found
\emph{c}-axis ZBCP,\cite{Yuli PRB} we further pursued our STS
study on (001)La$_{1.88}$Sr$_{0.12}$CuO$_4$ films in order to
examine in detail the spatial evolution of the LDOS and, in
particular, of the ZBCP. Our main finding is an anisotropy in the
spatial dependence of the ZBCP amplitude. Along lines of
sequentially acquired tunneling spectra we found a modulation of
the ZBCP amplitude with a modulation length larger than the 4$a_0$
inter-stripe spacing reported for LSCO(x=0.12).\cite{Yamada} At
the same rate of recurrence, lines of constant ZBCP amplitude were
measured on the same area but at different directions.
Occasionally, the modulated ZBCP was split into two imbalanced
peaks. Our diverse spectroscopic features comply well with the
anti-phase ordering of the superconductor order parameter and its
predicted effect on the LDOS \cite{Sigrist} in the limit of
negligible coupling between adjacent $\pi$DWs. In addition, our
results point to an inhomogeneous distribution of the pairing
amplitude at the sample surface.

\begin{figure}
\includegraphics[width=3.5in]{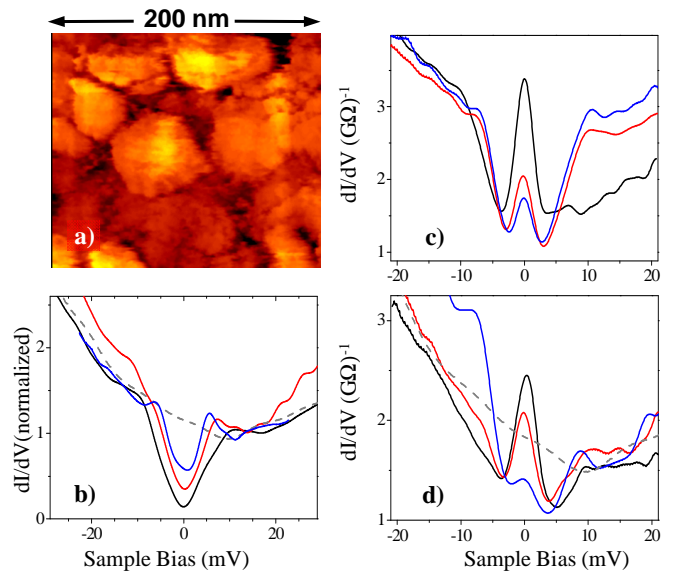}
\caption{(Color online) a) 200 x 200 nm$^2$ topographic image of
the sample surface showing LSCO crystallites. b-d) Typical
tunneling spectra of our \emph{c}-axis LSCO(x=0.12) samples. b)
V-shaped tunneling gaps expected to be found on \emph{c}-axis
surfaces, however these gaps consisted the minority of our data.
The spectra were normalized with respect to the normal tunneling
conductance at 13 mV, which exceeds the gap value. c) and d) The
ZBCP, which was the predominant spectral feature of our data. Note
that suppressed gap-like features are typically observed alongside
the pronounced ZBCPs. The dashed curves in b) and d) portray the
smearing effect of the surface disorder on the low-bias features.
This type of spectra denoted ends or interruptions of the spectra
lines.} \label{fig1}
\end{figure}

\section{EXPERIMENT}

Four 90 nm (001)La$_{1.88}$Sr$_{0.12}$CuO$_4$ films were
epitaxially grown using the pulsed laser deposition technique on
(100)SrTiO$_3$ substrates, as detailed elsewhere.\cite{Yuli PRB}
In addition, a nodal-surface oriented
(110)La$_{1.88}$Sr$_{0.12}$CuO$_4$ film was grown on
(110)SrTiO$_3$ for comparison reasons. Despite our efforts, we
were unable to obtain sharp topographic images (i.e. atomic-scale
resolution) of the sample's surface due to the notoriously
degraded surface of LSCO films \cite{Kato} and only correlations
of the spectra with the gross morphological features, namely the
crystalline structure of the films, were obtained, as demonstrated
in our previous report.\cite{Yuli PRB} A typical topographic map
showing LSCO crystallites is presented in Fig. 1(a), allowing us
to verify that the ZBCP features are not due to any side facet of
the \emph{c}-axis crystallite, and in particular not to (110)
facets. Although correlations with the fine, atomic-scale, surface
morphology of the samples could not be obtained, by acquiring our
tunneling spectra along intersecting perpendicular lines, the
spatial dependence of the surface DOS was elucidated to some
extent. The length of the lines along which the spectra was
acquired (at equidistant steps) was limited to a few tens of nm,
at which point the spectra usually turned featureless (probably
due to surface disorder), as demonstrated by the dashed curves in
Figs. 1(b) and 1(d). The spectra were acquired with
100M$\Omega$-1G$\Omega$ tunneling resistance.

\section{RESULTS and DISCUSSION}

\begin{figure}
\includegraphics[width=3.1in]{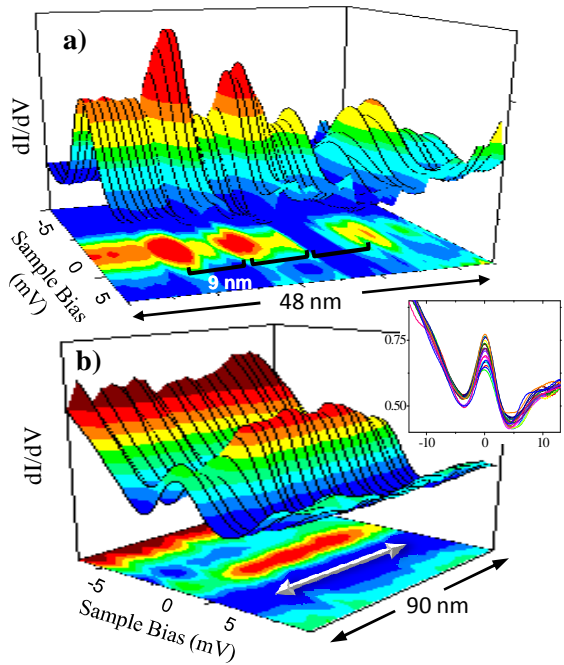}
\caption{(Color online) Two spectra-lines showing modulated ZBCP
amplitude. a) The shortest modulation length of 3-4 nm is seen for
all three ZBCP amplitude maxima. The inter-maxima distance is 9 nm
for the most pronounced ZBCPs. b) The longest modulation length we
encountered of $\sim$ 60 nm. Inset: The dI/dV vs. V curves taken
in the region marked by the double-headed arrow at the base of the
figure.} \label{fig2}
\end{figure}

Tunneling into a \emph{c}-axis surface of a \emph{d}-wave
superconductor yields V-shaped gaps due to the zero-energy nodal
excitations. Part of our tunneling spectra featured such a
V-shaped gap as presented in Fig. 1(b). Surprisingly, however,
these gaps constituted less than 10\% of the spectra that
exhibited clear superconductor features, which were found on
30-40\% of the sample surface. The majority of the tunneling
spectra exhibiting superconducting features, portrayed a ZBCP with
gap-like features (GLFs). The variance of the ZBCP amplitude is
shown in Figs. 1(c) and 1(d), which present two sets of dI/dV vs.
V curves (tunneling spectra), each taken along an individual line.
Interestingly, it was commonly found that a pronounced peak
entailed suppressed GLFs and vice versa [as evident from Figs.
1(c) and 1(d)], reminiscent of the predicted \cite{Sigrist}
spectral-weight shift as a function of distance from an isolated
$\pi$DW [see Fig. 1(a) in Ref. \onlinecite{Sigrist}]. However, no
clear correlation was found between the ZBCP and GLF for the
intermediate amplitudes, possibly due to the inhomogeneity of the
superconductor-gap, regularly found in the tunneling spectra of
cuprates.\cite{Kato, Davis, McElroy, Yazdani, Kapitulnik} One
should recall that the LDOS of a \emph{d}-wave superconductor
should feature a similar ZBCP with GLFs when tunneling into the
\emph{ab}-plane (excluding the anti-nodal direction). In such
setups, the GLF to ZBCP magnitude ratio is indicative of the
in-plane tunneling angle with respect to the nodal direction.
Obviously, this is not the case in our present \emph{c}-axis
films.

\begin{figure}
\includegraphics{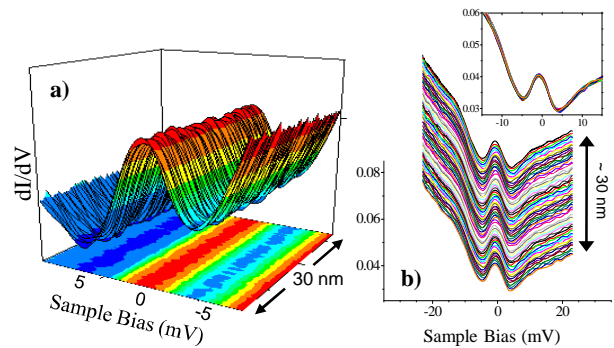}
\caption{(Color online) a) Two spectra-lines showing non-modulated
ZBCP amplitude. The scan line in a) was taken on a line
perpendicular and intersecting to that presented in Fig. 2(a). The
line in b) demonstrates that not only the ZBCP but also the GLF
remains constant for the entire energy range we examined, as
clearly portrayed in the inset where the same spectra plotted
without any shift, appear to collapse onto a single dI/dV vs. V
curve.} \label{fig3}
\end{figure}

Due to the absence of a clear correlation between the mid-gap and
gap-edge states we focus on the spatial evolution of the low
energy states, namely the evolution of the ZBCP. In Figs. 2 and 3
we plot the dI/dV vs V tunneling spectra as a function of position
of their acquisition. The detailed spatial evolution of the LDOS
exhibited two distinct types of spatial dependencies. The first
type consisted of a modulated ZBCP amplitude with a modulation
length ranging from a few nm to tens of nm. The shortest
modulation length we encountered is presented in Fig. 2(a) in
which three local maxima of the ZBCP amplitude develop, each
having a spatial width of 3-4 nm. About 9 nm part the two most
pronounced ZBCPs, and $\sim$18 nm separate the next pronounced
peak (of reduced height). In between, and just midway the latter
two peaks, one can see the effect of the surface disorder
mentioned above, smearing the ZBCPs on a short length scale. The
longest modulation scale we found was close to 60 nm and is
presented in Fig. 2(b), in which a single wide maximum of the ZBCP
amplitude was recorded along the line. The slow variation of the
ZBCP magnitude within the segment around the maximal height
(marked by a white double-headed arrow in the base of the figure)
is demonstrated in the inset. Towards the edges of this line scan,
a continuous reduction of the zero energy spectral weight is
noticeable over a length scale of 10-20 nm, an order of magnitude
larger than the decay length shown in Fig. 2(a).

The second type of spatial dependence consisted of a constant ZBCP
amplitude. Two typical examples are presented in Fig. 3, and
evidently, these curves show a minuscule variation of the LDOS
along the line. This is clearly illustrated in the inset of Fig.
3(b) where all the dI/dV curves collapse onto a single curve when
all the spectra obtained along the entire line are plotted.
Typically, the DOS in such cases was robustly fixed for the entire
spectrum we examined, and not only at low energies on which the
spectra in Fig. 3(a) focus. Fig. 3(b) demonstrates the robustness
of the LDOS for tunneling spectra taken along a 30 nm long line,
on a different sample. It is important to note that the lines in
Figs. 2(a) and 3(a) were measured sequentially at perpendicular
and intersecting trajectories.

\begin{figure}
\includegraphics[width=3.2in]{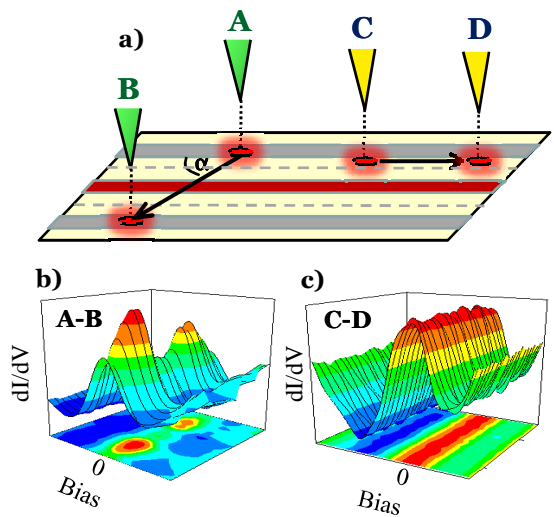}
\caption{(Color online) a) Schematic illustration of the
one-dimensional anti-phase order. The colored lines represent the
domain centers and the $\pi$DW is represented by the dashed line.
b) A constant ZBCP is expected for tunneling spectral lines in
parallel to the DW like the one connecting points C and D in (a).
c) A modulation of the ZBCP amplitude should occur for any line
with $\alpha \neq$ 0. In the case of the line connecting the
points A and B, a double peak structure will emerge since it
crosses two $\pi$DWs.} \label{fig4}
\end{figure}

The parallels of these results with the framework of the
aforementioned isolated $\pi$DW scenario,\cite{Sigrist} are
compelling. In the limit of decoupled $\pi$DWs, the LDOS were
predicted to consist of a ZBCP featuring a spatially dependent
amplitude determined by the distance from the DW. Consequently,
lines parallel to a $\pi$DW should exhibit a constant ZBCP
amplitude while any trajectory intersecting a DW at an angle
$\alpha$, will result in a modulated ZBCP amplitude with an
$\alpha$-dependent modulation length, corresponding to the
effective distance from the DW. Multiple decoupled $\pi$DWs may
induce multiple maxima in the ZBCP magnitude along a line crossing
them with spacing that depends on the inter-$\pi$DW distance and
$\alpha$. The sketch in Fig. 4(a) illustrates two different types
of lines within the anti-phase ordered plane along which tunneling
spectra can be acquired, one running parallel to the $\pi$DWs and
one crossing them. The spatial evolution of the LDOS in Figs. 4(b)
and 4(c) are segments taken from the line scans presented in Figs.
2(a) and 3(a), respectively. Both lines comply well with the two
different expected behaviors corresponding to the lines sketched
in 4(a). The line connecting A and B in Fig. 4(a) crosses two
$\pi$DWs (dashed lines), hence, a double-peak structure emerges in
the corresponding spectra acquired along such a line, as shown in
Fig. 4(b). In contrast, the parallel line connecting C and D
should produce a constant ZBCP line, as shown in Fig. 4(c). Note
that at a right angle ($\alpha = \pi/2$), the distance between the
most pronounced ZBCPs (inter-maxima distance) will be minimal and
equal to the local spacing between adjacent $\pi$DWs.

Following the above considerations, we attribute our results to a
state of decoupled $\pi$DWs by assigning an appropriate angle,
$\alpha$, to each line of spectra presented in Figs. 2 and 3.
Obviously, $\alpha$=0 for the lines exhibiting a constant ZBCP in
Fig. 3. A small angle will precipitate the slow variation in the
LDOS portrayed in Fig. 2(b), since the effective distance from the
DW varies slowly along the line. In contrast, the rapid modulation
of the LDOS in Fig. 2(a) will occur at large angles. The fact that
the the spectra presented in Fig. 2(a) were measured along a line
intersecting at a right angle the line at which the spectra
presented in Fig. 3(a) were acquired, indicates that for Fig. 2(a)
$\alpha=\pi$/2, which implies that the observed 9 nm inter-maxima
interval corresponds to the local distance between the neighboring
$\pi$DWs. Such distances are consistent with the assumed
negligible $\pi$DW coupling, since they are almost twice the
spatial width of the ABS, which is of the order of $\xi_s$ ($\sim$
5 nm in the case of LSCO). We also note that the modulation scale
is larger than the spacing between charge stripes ($\sim$ 1.5 nm),
an issue that is discussed further below.

\begin{figure}
\includegraphics[width=3in]{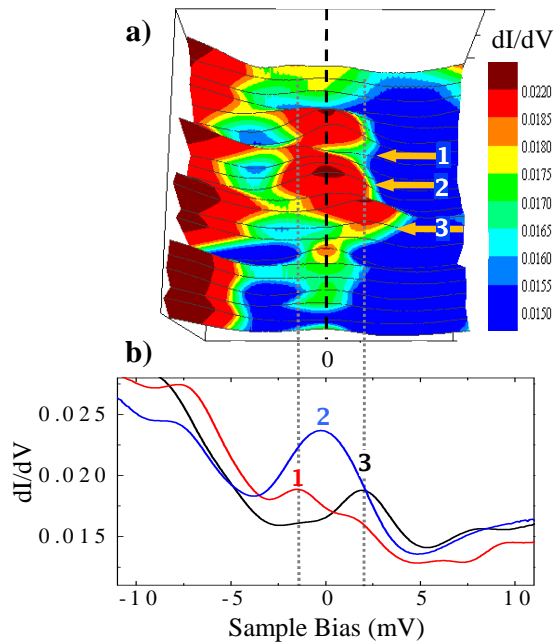}
\caption{(Color online) a) The spatial evolution of a split ZBCP
corresponding to a $\pi$DW at point 2 in the case of strong local
pairing (see text). b) dI/dV vs. V curves exhibiting split ZBCPs
and a pronounced centered (un-split) ZBCP taken at the indicated
positions in (a).} \label{fig5}
\end{figure}

A subtle feature of the modulated ZBCP, appearing only
occasionally in the spectra lines, was an asymmetric splitting
with a varying degree of imbalance between the negative and
positive peak heights. The imbalance ranged from a small
difference in peak heights to a state where one of the peaks was
completely suppressed. The transition from a nearly fully
suppressed positive peak, to a nearly fully suppressed negative
peak, through a centered (un-split) ZBCP, could take place even
within a distance of a few nm, as demonstrated in Fig. 5(a) and
highlighted by the three spectra presented in Fig. 5(b).
Interestingly, the imbalance of the peaks in Fig. 5(a) meandered
continuously between the points indicated by the arrows labelled 1
and 3, and in between (at point 2), where a pronounced un-split
ZBCP was observed, the low energy spectral weight reached a maxima
as clearly depicted in Fig. 5(b). Also note that dI/dV vs. V
curves featuring a split ZBCP [i.e. the spectra in the region
enclosed by arrows 1 and 3 in Fig. 5(a)] exhibited a relatively
high spectral weight at low energies compared to the curves at the
bottom and top ends of the line, where a suppressed un-split ZBCP
was found.

A markedly different spatial evolution of a split ZBCP was
apparent along the line depicted in Fig. 6(a). Here, a periodic
modulation was measured with a period of $\sim$4 nm. Unlike the
line in Fig. 5(a), the imbalance did not cross the zero energy.
Instead, the negative peak was consistently higher than the
positive peak throughout the entire line, as shown in Fig. 6(b),
in which a selection of dI/dV vs V curves from Fig. 6(a) are
plotted.

According to the decoupled $\pi$DW scenario \cite{Sigrist} (as
mentioned in Sec. \ref{sec:intro}), in the case of a moderate
pairing strength, an imbalanced split of the ZBCP is expected to
occur at the vicinity of the DW. At sufficiently long distances
from the DW the splitting disappears and a suppressed ZBCP is
predicted to be found, as in the case of the weak pairing limit,
as indeed shown in Fig. 5. Consequently, the spatially modulated
split ZBCPs presented in Figs. 5 and 6, can be understood within
the framework of the decoupled $\pi$DW scenario assuming a locally
moderate pairing strength (as opposed to the data presented in
Figs. 2 and 3, where the data complies with a relatively weak
pairing strength). We note that lines of spectra showing either
split or un-split ZBCPs were found on the same sample, suggesting
(according to Ref. \onlinecite{Sigrist}) that the local pairing
amplitude in our samples is spatially inhomogeneous. This is a
reasonable conjecture in view of the superconducting-gap
inhomogeneities reported in the cuprates \cite{Davis} which have
been considered to reflect an inhomogeneity of the local pairing
strength.

Typically, the inversion of the split peak imbalance should take
place at the $\pi$DW owing to the sign change of the order
parameter. This implies that a $\pi$DW resides between points 1
and 3 in Fig. 5(a). The suppressed un-split ZBCP measured at the
bottom and top ends of Fig. 5(a) further corroborates this
conjecture, since according to the isolated $\pi$DW scenario, the
splitting occurs only in the vicinity of the DW where, in
parallel, the low-energy spectral weight is maximal.

\begin{figure}
\includegraphics[width=3in]{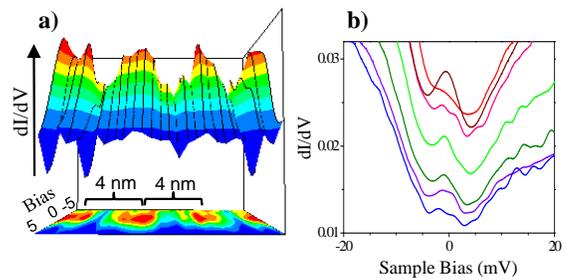}
\caption{(Color online) a) A modulated split ZBCP with a very
short modulation period. b) Selected dI/dV vs. V curves taken
along the line in (a). The ZBCP appeared to be split with a
negative peak consistently higher than the positive peak
suggesting that no $\pi$DW has been crossed. All curves are
vertically shifted for clarity.} \label{fig6}
\end{figure}

We now turn to discuss the spectra-line presented in Fig. 6 that
exhibits a periodic modulation of the negative-bias peak height of
the split ZBCP, with a period length of $\sim$ 4 nm, close to
$\xi_s$. Such a short period seems to contradict the isolated
$\pi$DW scenario since at these distances the inevitable overlap
of the neighboring ABSs is predicted to quench the
ZBCP.\cite{Sigrist, Shirit} However, the fact that the negative
peak was consistently higher than the positive peak indicates that
a $\pi$DW was not crossed and the line in Fig. 6(a) is in fact
parallel to a $\pi$DW. In this case it is possible that the
observed modulation is an outcome of the conjectured
\cite{Sigrist} one-dimensional band formed along the $\pi$DW in
which the wavefunction of a charge carrier should oscillate with a
period corresponding to the spatial width of an ABS, namely $\sim
\xi_s$.

We have also measured the tunneling spectra of a
(110)La$_{1.88}$Sr$_{0.12}$CuO$_4$ reference film where ZBCP
manifesting the nodal ABSs are the common spectral feature. In
this sample only weak spurious variations of the ZBCP and gap-like
features amplitude were found, in contrast to the well defined
spatial modulations seen on the (001)La$_{1.88}$Sr$_{0.12}$CuO$_4$
film, presented in Fig. 2. Moreover, no split ZBCPs were detected,
thus ruling out the possibility of spontaneous splitting due to a
subdominant order parameter as has been reported to occur in
overdoped (110)\YBCO \cite{DaganPRL,SharoniBTRS}.

A question that remains is whether the DWs are truly decoupled.
Theoretical studies of the anti-phase order predicted that
$\pi$-phase (and consequently $\pi$DW) ordering could be unstable
since the introduction of additional mid-gap states increases the
energy of the system.\cite{Sigrist, Scalapino} However, Yang
\emph{et al.} \cite{Sigrist} find that the energy cost decreases
as the domain size increases from the stripe periodicity of  4
unit cells, to 8 and 10 unit cells, although it remained positive
for all studied values. Thus according to Ref.
\onlinecite{Sigrist} it is more likely that an order of decoupled
anti-phase DWs (due to the large domain size) will form rather
than a system of overlapping $\pi$DWs once an anti-phase ordering
sets in. We wish to emphasize however, that the $\sim$ 9 nm
spacing (corresponding to 6 stripe unit cells, 24a$_0$, in LSCO)
between ZBCP maxima observed for a line taken perpendicular to the
$\pi$DW does not imply that this indeed is the generic spacing
between $\pi$DWs. It may well be that, due to surface disorder,
some $\pi$DWs are quenched, yielding spatial variations in the
apparent $\pi$DW spacing.

Finally, we note that the theoretical works that analyzed arrays
of coupled $\pi$DWs  predicted that a gap will open at the
chemical potential rather than a ZBCP. In our previous report
\cite{Yuli PRB} we have already disclosed that $\sim$ 10\% of the
superconductor related data comprised a V-shaped gap. A gap of
$\sim$ 10 meV was extracted by fitting to the theory of tunneling
into a \emph{d}-wave superconductor.\cite{TK} It is possible that
this gap is truly the \emph{c}-axis pairing gap found at large
enough distances from a $\pi$DW so that its effect is negligible.
On the other hand, it may well be that the gap is not directly
related to the superconducting order parameter and rather
originates from the local overlap of a neighboring $\pi$DWs. This
notion gains support in light of the relatively large variations
in the gap features (e.g. width and zero-bias conductance) shown
in Fig. 1(b).

\section{Summary and conclusions}

Our scanning tunneling spectroscopy data measured on \emph{c}-axis
La$_{1.88}$Sr$_{0.12}$CuO$_4$ films exhibit various features that
are in accord with the predicted LDOS of the anti-phase ordering,
in the limit of decoupled domain walls. The abundance of the ZBCPs
found on our \emph{c}-axis films is attributed to the formation of
ABSs at the $\pi$DW and their spatial evolution is a consequence
of the distance of the location of measurement from the nearby
$\pi$DW. The spatial evolution of the imbalanced split ZBCPs,
which is another trademark of the $\pi$DW model, points to an
inhomogeneity of the pairing strength within the sample surface.
We conclude that our results are indicative of a decoupled
$\pi$DWs state that provides possible evidence for the predicted
anti-phase ordering of the superconductor order parameter at x =
1/8. To the best of our knowledge this is the first experimental
indication for an anisotropic ordering of the \emph{d}-wave order
parameter.

\section{Acknowledgments}

The authors are grateful to S. Baruch, A. Frydman, D. Orgad, S.
Kivelson and M. Sigrist for stimulating discussions. This research
was supported by the Israel Science Foundation, Center of
Excellence Program (grant No. 481/07), the United States - Israel
Binational Science Foundation (grant No. 2008085), the Harry de
Jur Chair in Applied Science, the Heinrich Hertz Minerva Center
for HTSC, the Karl Stoll Chair in advanced materials and by the
Fund for the Promotion of Research at the Technion.


\begin{thebibliography}{999}

\bibitem {Matsuzaki}
T. Matsuzaki, M. Ido, N. Momono, R.M. Dipasupil, T. Nagata, A.
Sakai and M. Oda, J. Phys. Chem. Sol. \textbf{62}, 29 (2001).

\bibitem {Sato2}
H. Sato, A. Tsukada and M. Naito, Physica C \textbf{408-410}, 848
(2004).

\bibitem {Mood}
A. R. Moodenbaugh, Y. Xu, M. Suenaga, T. J. Folkerts and R. N.
Shelton, Phys. Rev. B \textbf{38}, 4596 (1988).

\bibitem{Yamada}
K. Yamada, C. H. Lee, K. Kurahashi, J. Wada, S. Wakimoto, S. Ueki,
H. Kimura, Y. Endoh, S. Hosoya, G. Shirane, R. J. Birgeneau, M.
Greven, M. A. Kastner and Y. J. Kim, Phys. Rev. B \textbf{57},
6165 (1998).

\bibitem{Tran}
J. M. Tranquada, B. J. Sternlieb, J. D. Axe, Y. Nakamura, and S.
Uchida, Nature \textbf{375}, 561 (1995).

\bibitem{Himeda}
A. Himeda, T. Kato and M. Ogata, Phys. Rev. Lett. \textbf{88},
117001 (2002).

\bibitem{Orgad PRB}
D. Orgad, Phys. Rev. B \textbf{79}, 014509 (2009).

\bibitem {Li}
Q. Li, M. H\"{u}cker, G. D. Gu, A. M. Tsvelik and J. M. Tranquada,
Phys. Rev. Lett. \textbf{99}, 067001 (2007).

\bibitem{B}
V. L. Berenzinskii, Zh. Eskp. Teor. Fiz. \textbf{61}, 1144 (1971)
[Sov. Phys. JETP \textbf{34}, 610 (1972)].

\bibitem{KT}
J. M. Kosterlitz and D. J. Thouless, J. Phys. C \textbf{6}, 1181
(1973).

\bibitem {He}
R.-H. He, K. Tanaka, S.-K Mo, T. Sasagawa, M. Fujita, T. Adachi,
N. Mannella, K. Yamada, Y. Koike, Z. Hussain and Z.-X Shen, Nat.
Phys. \textbf{5}, 119 (2009).


\bibitem {Berg}
E. Berg, E. Fradkin, E.-A. Kim, S. A. Kivelson, V. Oganesyan, J.
M. Tranquada and S. C. Zhang, Phys. Rev. Lett. \textbf{99}, 127003
(2007).

\bibitem{TK LDOS}
Y. Tanaka and S. Kashiwaya, Phys. Rev. B \textbf{53}, 9371 (1996).

\bibitem{Sigrist}
K.-Y. Yang, W. Q. Chen., T. M. Rice, M. Sigrist and F.-C. Zhang,
New J. Phys. \textbf{11}, 055053 (2009).

\bibitem {Shirit}
S. Baruch and D. Orgad, Phys. Rev. B \textbf{77}, 174502 (2008).

\bibitem {Yuli PRB}
O. Yuli, I. Asulin, O. Millo and G. Koren, Phys. Rev. B
\textbf{75}, 184521 (2007).

\bibitem {Hu}
C.-R. Hu, Phys. Rev. Lett. \textbf{72}, 1526 (1994).

\bibitem {TK}
S. Kashiwaya, Y. Tanaka, M. Koyanagi and K. Kajimura, Phys. Rev. B
\textbf{53}, 2667 (1996).

\bibitem {Dagan}
Y. Dagan, A. Kohen, G. Deutscher and A. Revcolevschi, Phys. Rev. B
\textbf{61}, 7012 (2000).

\bibitem {Greene2} H. Aubin, L. H.
Greene, S. Jian and D.G. Hinks, Phys. Rev. Lett. \textbf{89},
177001 (2002).

\bibitem {Kato}
T. Kato, S. Okitsu and H. Sakata, Phys. Rev. B \textbf{72}, 144518
(2005).

\bibitem {Davis}
K. M. Lang, V. Madhavan, J. E. Hoffman, E. W. Hudson, H. Eisaki,
S. Uchida and J. C. Davis, Nature, \textbf{415}, 24 (2002).

\bibitem {McElroy}
J. E. Hoffman, K. McElroy, D.-H. Lee, K. M Lang, H. Eisaki, S.
Uchida and J. C. Davis, Science, \textbf{297}, 1148 (2002).

\bibitem {Yazdani}
M. Vershinin, S. Misra, S. Ono, Y. Abe, Y. Ando and A. Yazdani1,
Science, \textbf{303}, 1995 (2004).

\bibitem {Kapitulnik}
C. Howald, H. Eisaki, N. Kaneko, M. Greven and A. Kapitulnik,
Phys. Rev. B \textbf{67}, 014533 (2003).

\bibitem{DaganPRL}
Y. Dagan and G. Deutscher, Phys. Rev. Lett. \textbf{87}, 177004
(2001).

\bibitem{SharoniBTRS}
A. Sharoni, O. Millo, A. Kohen, Y. Dagan, R. Beck, G. Deutscher
and G. Koren, Phys. Rev. B \textbf{65}, 134526 (2002).

\bibitem {Scalapino}
S. R. White and D. Scalapino, arXiv:0810.0523 (2008).




\end{thebibliography}
\end{document}